# Making Silicon Emit Light Using Third Harmonic Generation

Abdurrahman Javid Shaikh[a*], Othman Sidek[a]

*Collaborative µElectronic Design Excellence Centre (CEDEC), Engineering Campus,
Universiti Sains Malaysia, 14300 Nibong Tebal, Seberang Perai Selatan, Pulau Pinang, Malaysia.*

**Abstract**

Despite its excellent performance in microelectronic industry, silicon was not able to perform well in photonic devices arena. This is because the silicon has never been a good optical source mainly due to its indirect band gap structure. Many of the device functionalities in silicon have been reported, with an exception of – until recently – a reliable optical source. Silicon is a nonlinear material which makes use of its nonlinearities to realize various functionalities. This paper presents a theoretical treatment of generating and enhancing third-harmonic field which may be used as optical source, crystal state monitoring and all-optical signal processing applications.

## 1. Introduction

Lasing on silicon substrates has been of profound interest for scientists since successful demonstration of silicon based transistor radios couple of years after birth of the first laser. But it was not until 1980s when Soref (et al.) initially suggested the use of silicon crystal as material for all-optical components that brought an inclined interest towards silicon photonics research [1, 2]. With continued reduction in device size and inception of multi-core microprocessors was created a bottle-neck at processor-interconnect interface primarily due to the speed limitation of electrical wires. This constraint suggested to resort to alternative media having virtually infinite bandwidths and distance independence, that is, optical interconnects [3, 4].

---

* Corresponding author. Tel.: +60-13-5072908
*E-mail address*: abdurrahman.j.shaikh@gmail.com .

The backbone of any photonic component and photonic integrated circuit (PIC), whether active or passive, is a waveguide which acts as field transformation medium for optical signals. Silicon's most attractive strength is the ability to get ultra pure waveguide out of it at very cheap costs [5]. Material properties of silicon, its compatibility with CMOS VLSI technology together with successful demonstration of passive and active functionalities, brought an impetus in and a major focus shift of the research community across the globe towards the silicon based photonic devices. The primary objective of this focus is the ability to get all the functionalities "optically", in contrast with "electronically", on a single chip thereby ultimately reducing the cost as well as improving design & performance of a particular application [3]. Hence, silicon is now becoming the top-notch material for prospective photonics industry.

## 2. Theory of nonlinearity

As mentioned in the opening lines, silicon is a nonlinear material. The interaction of electric component of optical field with electrons and phonons inside silicon waveguide (WG) triggers the nonlinear phenomena. It can all be attributed to the polarization (**P**) of dielectric molecules / constituent atoms, by the applied electric field (**E**). The relationship of Eq. (1) following, can be found in many textbooks dealing with nonlinear optics.

$$P = \varepsilon_0 (|\chi^{(1)}||E| + |\chi^{(2)}||E|^2 + |\chi^{(3)}||E|^3 + ...) \qquad (1)$$

where, $\varepsilon_0$ is the permittivity of free space and $\chi^{(1)}$ is the linear first-order optical susceptibility, while $\chi^{(2)}$ and $\chi^{(3)}$ are the nonlinear second and third order susceptibilities and so on. For electric field intensities of smaller magnitude, the higher order susceptibility terms of (1) are negligible, nonetheless, they become prominent as the intensity gets significant [6]. However, the second order nonlinear susceptibility is absent in silicon because of its centrosymmetric structure. Hence, nonlinear processes initiated by third-order nonlinear susceptibility are prevalent in silicon, as excitation intensity becomes considerably large, where the nonlinear polarization is related to susceptibility by [7];

$$P = 4 |\chi^{(3)}| |E|^3 \qquad (2)$$

The third order nonlinear susceptibility $\chi^{(3)}$ gives rise to nonlinear phenomena like *stimulated Raman scattering (SRS)*, *self phase modulation (SPM), four-wave mixing (FWM), cross phase modulation (XPM)* and *third harmonic generation (THG)*. Hence the nonlinear susceptibilities, when plugged into Eq. (1), result in intensity dependent nonlinear refractive index change culminating in highly nonlinear response of silicon to applied fields.

Because of its indirect band gap structure, silicon does not support efficient light emission by electron-hole recombination process. What dominate the recombination in silicon are non-radiative processes known as Free Carrier Absorption (FCA) and Auger recombination [8]. These non-radiative processes in the indirect band gap Si result in the quantum efficiency of the order $10^{-4}$%. On the other hand, two photon absorption (TPA) – a detrimental nonlinear process – results in pulling out of an electron from valence band to conduction band with the assistance of two photons absorbed in the crystal. Obviously, TPA increases free charge concentration in silicon crystal which, in turn, increases the probability of FCA. Hence, FCA, Auger recombinations and the TPA make light emission from silicon highly unlikely.

TPA poses another serious problem to photonic devices based on silicon which can be described as follows. Change in refractive index due to applied optical field intensity is referred to as *optical Kerr effect*. This effect in silicon, in spite of having appreciable optical Kerr coefficient value *(*$n_2$ = 4.0 x 10$^{-18}$ m$^2$/W [9] at 1.54 μm*)*, is very weak mainly due to comparatively large TPA coefficient (β = 8 x 10$^{-12}$ [9] at 1.54 μm) near telecommunication wavelengths (i.e. 1.31 μm & 1.55 μm). These values lead to very low nonlinear optical figure of merit (FOM) – for example – lesser than 0.4 at 1.54 μm. However, an FOM greater than 2 is desirable to observe significant optical Kerr nonlinearities in a third-order nonlinear medium (Kerr medium o Kerr cell) like silicon.

From Fig. 1 [10], TPA can be observed to have negligible values at photon energies lesser than half the indirect band gap energy of silicon ($E_{ig}$ = 1.12 eV) – that is at wavelengths beyond 2.2 μm. On the other hand, the optical Kerr coefficient has substantial value at wavelengths greater than ~2.2 μm. Therefore, the figure of merit of silicon is much improved in the mid-IR region, hence silicon has better prospects to serve as a gain medium. Optical sources within this spectrum have prospective applications in medicine and clinical surgery as these frequencies have maximum interaction with human tissue.

The optical Kerr effect, arising from third order nonlinear susceptibility ($\chi^{(3)}$) which itself becomes substantial only at significantly high field intensities, gives rise to intensity dependent nonlinear processes, one of which is the THG. This process has been utilized to demonstrate visible light emission from silicon photonic crystal (PhC) WG **[11]**. Following we treat this nonlinear process and discuss the experiment which demonstrated light emission from silicon.

## 3. Third-harmonic generation in silicon

THG is a third order Kerr nonlinearity in which three photons of same frequency (ω) from pump are absorbed by a waveguide / nanowire crystal and a new single photon of thrice the frequency (3ω) is emitted. Replacing applied monochromatic field intensity, |E|, by its equivalent (Re{|E(ω)|e$^{jωt}$}) in Eq. (2) gives third-harmonic component of the polarization field (indicating THG), as well as the fundamental component as follows [7];

$$P(\omega) = 3\chi^{(3)}|E(\omega)|^2 E(\omega) \qquad (3)$$
$$P(3\omega) = \chi^{(3)}|E(\omega)|^3 \qquad (4)$$

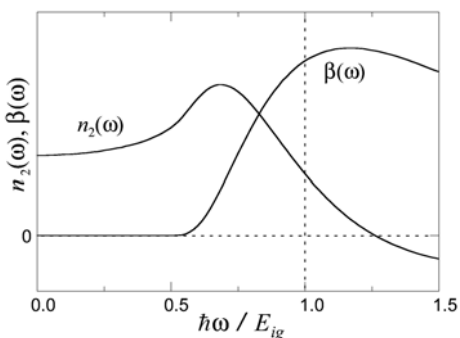

Fig. 1: Optical Kerr Coefficient ($n_2$) and Two Photon Absorption (β) vs normalized Photon Energy (horizontal axis). Copyright © 2003 IEEE. All rights reserved. Reprinted, with permission, from [10].

Conversion efficiency of THG from the applied fundamental field in terms of energy is quite low, hence, its contribution is usually least among the other nonlinear processes inside a Kerr cell [12]. Generally, in many cases third-harmonic (TH) is obtained by second-harmonic (SH) generation and then sum-frequency generation of the fundamental and SH fields [7]. However, in centrosymmetric crystals like silicon and germanium, SH does not exist as it requires absence of inversion symmetry [13]. On the other hand, THG does not require this condition. In fact, interference among TH waves generated in a sample can cause a detectable THG signal [14]. Moreover, THG intensity has been shown to demonstrate change with respect to dopant dose (*dose dependency*) in a sample and crystal rotation about axis normal to the sample (*rotational dependency*) for a particular growth orientation ((100), (110) and (111)) [13]. The intensity was found to have decreased with increasing dose until a threshold was reached after which it raised and achieved a relatively lower constant value. Dopants such as Boron, Phosphorus and Arsenic were used and the threshold was observed to decrease as dopant mass increased. The behaviour of silicon above the threshold dose suggested formation of amorphous layer inside silicon sample thereby less THG intensity which is equally true for silicon as a dopant in the sample. Hence, any involvement of impurity absorption is negligible [13].

Like any nonlinearity, THG in silicon can be enhanced (i) by making the optical field travel slow inside the WGs (slow light) [15], and (ii) by increasing optical field confinement [16]. The former increases probability of photon interaction with WG crystals while later enhances the electric field inside the WG. Increased electric field inside a WG means increased energy density which, in turn, means occurrence of nonlinear phenomena at much reduced input power (i.e., low-threshold). Whether it is high optical field confinement or achieving slow light, large index contrast and large refractive index of WG material are keys to achieve them. The high optical field confinement within a WG is understood due to large step between indices of silicon core and insulator ($SiO_2$) cladding ($\Delta n \sim 2.1$). Besides, tight optical field confinement allows fabrication of WGs with ultra-compact bending of radius in micron range [17] hence making the device CMOS-compatible. A tremendously improved solution to enhance slow light effect is to use photonic crystal silicon waveguide which also offers tight optical field confinement. Fig. 2 [15], shows a typical scanning electron microscope (SEM) of PhC Si WG.

THG has been utilized in frequency up-conversion to make an emissive source at green light frequency. Concoran, et al, in [11], utilized slow-light to enhanced THG and demonstrated directional green light emission from a silicon PhC waveguide. A two dimensional (2D) PhC WG of silicon – shown in Fig. 3(a) [11] – was pumped by near-IR pulses of 1.5ps with repetition rate of 4MHz. The work reported maximum emitted green light power at the pump wavelength of ~1557 nm. The wavelength of the emission was verified to be 520 ± 5 nm (~wavelength of green light) by two band pass filters. The maximum THG output power of ~10 pW was observed for average pump power of ~80μW. By virtue of the high field confinement, peak power coupled into the waveguide was around 10W which is 5 to 6 order of magnitude smaller than the power needed to be coupled in bulk silicon. Also, increasing the electric field intensity of the fundamental light increases THG power [16]. Since silicon rapidly absorbs visible light, the green light emitted must be readily removed out of the waveguide. Apart from slow light enhancement, the photonic crystals allowed a fraction of emissive mode at 3ω to be coupled out perpendicular to the opaque silicon waveguide plane. The SEM image of the silicon PhC WG, as reported in [11], is given in Fig. 3(b) (labeled).

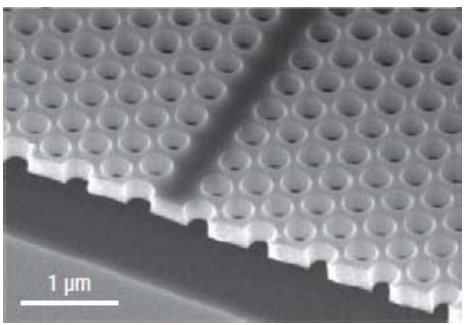

Fig. 2: Scanning Electron Microscope (SEM) image of a silicon photonic crystal waveguide. Reprinted by permission from Macmillan Publishers Ltd: Nature Photonics [15], copyright 2008.

## 4. Applications

Optical third harmonic generation has been successfully demonstrated for all-optical performance monitoring of optical signals, based on silicon, with processing rate up to 640 Gbit/s [18]. Since implanted ion dose affects the THG intensity implying change in crystalline state, optical THG can also be utilized as a means of monitoring state transition of silicon (from crystalline to amorphous) [13].

## 5. Conclusion

In this paper, we discussed in detail third harmonic generation in silicon, factors affecting it and detailed the aspects of published work. FCA, Auger recombinations due to indirect band gap structure of silicon, and TPA in the telecommunication wavelength regime posed serious problems to realizations of photonic integrated circuits at these technologically important frequencies. Nonetheless, green light emission at the peak 10 pW has been successfully reported using slow light which spatially compressed the optical field and dramatically enhanced nonlinearity in silicon. This report is a significant achievement as far as using silicon as a light source is concerned, however, it may not be considered as a replacement of comparatively high power and highly collimated silicon based lasers which are required for on-chip / intra-chip optical communication. Still, this report has given hope to keep seeking optical sources based on silicon as until recent past it was considered an impossible task.

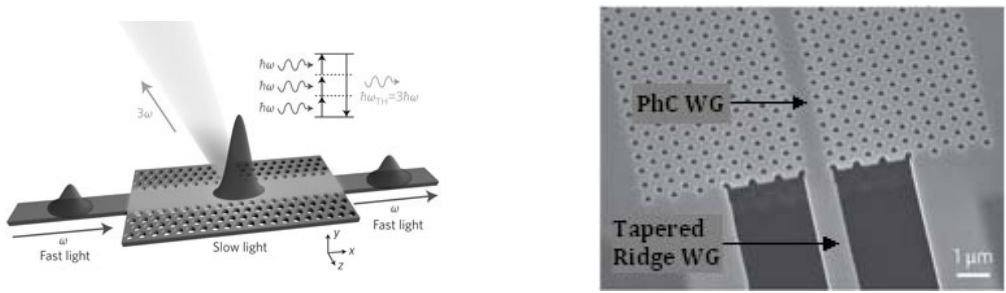

Fig. 3. (a) "Schematic of slow-light enhanced THG. The fundamental pulse at frequency ω (energy ℏω) is spatially compressed in the slow-light photonic-crystal waveguide, increasing the electric field intensity, while the third-harmonic signal, at frequency $\omega_{TH} = 3\omega$, is extracted out-of-plane by the photonic crystal with a specific angle off the vertical direction."; (b) "Scanning electron microscope (SEM) image of a silicon PhC waveguide connected to the tapered ridge access waveguide." Reprinted by permission from Macmillan Publishers Ltd: Nature Photonics [11], copyright 2009.